# Unsupervised super-spatial-resolution Brillouin frequency shift extraction based on physical enhanced spatial resolution neural network

**ZHAO GE,[1] HAO WU,[1,*] ZHIYONG ZHAO,[1] LI SHEN, AND MING TANG[1]**

*[1] Wuhan National Laboratory for Opto-electronics, Next Generation Internet Access National Engineering Laboratory, and Hubei Optics Valley Laboratory, School of Optics and Electronic Information, Huazhong University of Science and Technology, Wuhan, China, 430074*
** wuhaoboom@hust.edu.cn*

**Abstract:** Spatial resolution (SR), a core parameter of Brillouin optical time-domain analysis (BOTDA) sensors, determines the minimum fiber length over which physical perturbations can be accurately detected. However, the phonon lifetime in the fiber imposes an inherent limit on the SR, making sub-meter-level SR challenging in high-SR monitoring scenarios. Conventional SR enhancement approaches, constrained by hardware limitations, often involve complex systems, or increased measurement times. Although traditional deconvolution methods can mitigate hardware constraints, they suffer from distortion due to the nonlinear nature of the BOTDA response. Supervised deep learning approaches have recently emerged as an alternative, offering faster and more accurate post-processing through data-driven models. However, the need for extensive labeled data and the lack of physical priors lead to high computational costs and limited generalization. To overcome these challenges, we propose an unsupervised deep learning deconvolution framework, Physics-enhanced SR deep neural network (PSRN) guided by an approximate convolution model of the Brillouin gain spectrum (BGS). We validate PSRN on both simulated and experimental data. The results demonstrate that PSRN can reconstruct sub-meter SR BGS and accurately retrieve the corresponding Brillouin frequency shift (BFS) from any low-resolution BGS input in a plug-and-play fashion, leveraging the interplay between neural network inference and embedded physical priors. In the case of a 0.5 m hot-spot, the BFS retrieved by PSRN is highly consistent with that obtained from a supervised trained neural network (STNN). Unlike the STNN, our unsupervised method does not require labeled data or training process. Furthermore, our framework can solve the inverse problem with more general applicability, enabling high-SR BGS reconstruction and BFS retrieval across varying pulse widths, fiber lengths, and frequency sweep steps. This plug and play post-processing technique paves the way to enable novel high-SR BOTDA sensors, representing a significant advancement for Brillouin sensing applications

## 1. Introduction

Brillouin optical time-domain analysis (BOTDA) has emerged as a transformative technology in distributed fiber-optic sensing, distinguished by its unique capability to simultaneously measure temperature and strain variations along tens of kilometers of optical fiber with submeter spatial resolution (SR). Unlike conventional point sensors, BOTDA exploits the inherent physical interaction between propagating light and acoustic phonons in optical fibers, a phenomenon known as stimulated Brillouin scattering (SBS). By analyzing the Brillouin frequency shift (BFS) induced by localized temperature or strain changes, BOTDA enables high-sensitivity, high-precision monitoring over extended distances [1-3]. This technology has become indispensable for health monitoring of critical infrastructure, including oil and gas pipelines, power cables, and civil engineering structures [4-7]. However, conventional BOTDA systems face a fundamental limitation: the SR typically exceeds 1 m due to constraints imposed by the phonon lifetime (~10 ns). This restriction hinders accurate detection of sub-meter-scale localized temperature/strain variations, thereby limiting its applicability in high-SR scenarios such as aerospace structural health monitoring or mine safety assessments.

To address SR limitations in traditional BOTDA systems, the differential pulse-width pair (DPP) technique has been widely adopted [8-12]. In DPP-BOTDA, two Brillouin time-domain traces are independently acquired using long pump pulses with a slight width difference, and high-SR sensing signals are then extracted through trace subtraction. While this method offers advantages in implementation simplicity and long sensing distance, it necessitates doubled measurement time and exhibits heightened susceptibility to polarization fading noise and system instability [13]. Alternative approaches leverage signal post-processing algorithms to enhance SR [13-16]. For instance, by approximating Brillouin time-domain traces as a linear convolution between the pump pulse shape and the fiber's impulse response, deconvolution algorithms have achieved an SR of 0.2 m using 40 ns pump pulse [13]. However, because of the inertial features of the acoustic wave, BOTDA sensors cannot be rigorously regarded as a linear time-invariant system [13,17]. The Brillouin gain spectra (BGS) envelope becomes dependent on the detuned frequency along the fiber, violating the linear convolution assumption. This nonlinearity leads to notable distortions in the recovered results [13], particularly in regions with sharp and large BFS change. Although some sophisticated preprocessing methods have been proposed to eliminate this distortion, e.g., by using pulse differential preprocessing, it will however cause an increasement of measurement time [17,18].

Recently, deep learning (DL) has emerged as a revolutionary paradigm for solving inverse problems in scientific sensing [19,20], offering unique advantages in handling nonlinear mappings. A supervised learning-based SR enhancement method has been proposed [21], where models are trained on extensive simulated datasets to establish mapping relationships between BGS and their corresponding BFS. This approach has successfully demonstrated 0.5 m SR extraction from BGS measurements acquired with 40 ns pump pulses. However, training such models demands large-scale labeled datasets and computationally intensive procedures, rendering it time-consuming and energy-inefficient. Furthermore, the learned mapping functions are inherently pulse-width-specific, any change in pump pulse parameter and sweep frequency step necessitates model retraining, severely limiting practical adaptability.

In this paper, to overcome these limitations, we design and propose a novel SR improvement method based on unsupervised learning. We combine the convolutional model of BGS with a neural network, which we name physics-enhanced SR neural network (PSRN). PSRN does not require thousands of labeled data for training. Instead, it only needs a set of BGS input. Through self-supervised learning, the interaction between the neural network and the physical model optimizes the network's weights and biases, gradually refining them and ultimately achieving sub-meter SR improvement. Our method was validated through both simulation and experimental. PSRN accurately reconstructs high-SR BGS from low-SR BGS acquired under various pulse widths and frequency sweep steps. For experimental data with a 40 ns pulse width, the BFS retrieved by PSRN is highly consistent with that of a supervised trained neural network (STNN) method and significantly outperforms the conventional DPP method. Crucially, experimental results confirm that PSRN can be directly applied to BGS obtained under different pulse widths and sweep steps without any retraining, demonstrating true plug-and-play capability across diverse BOTDA measurement scenarios. This level of flexibility far surpasses that of supervised learning methods. To the best of our knowledge, this is the first application of physics-enhanced deep learning in the field of distributed optical fiber sensing. Given its breakthrough performance in overcoming the fundamental SR limits, this work holds substantial promise for the development of next-generation distributed sensing systems.

## 2. Methods

Fig. 1 illustrates the architecture of the proposed unsupervised PSRN for BFS extraction. The method requires only a set of raw BGS signal $y$ obtained from a single measurement using conventional BOTDA system. Notably, this approach dispenses with extensive labeled data and imposes no specific constraints on pulse width, frequency scanning step size, or BGS length. PSRN takes the low SR BGS measurement $y$ as its sole input. Through its manually designed structure, the network generates estimated parameters matrix $y_{pre} \in R^{3\times N}$, which includes the predicted BFS $v_B$, Brillouin linewidth $\Delta v_B$, and Brillouin gain coefficient $g_B$ along the fiber. Unlike supervised methods requiring true $v_B/\Delta v_B/g_B$ (often experimentally inaccessible) to optimize weights and biases by minimizing errors between $y_{pre}$ and labels. We compute the convolved BGS $\hat{y} = P(y_{pre})$, where P(·) denotes the convolution between the pump pulse and the fiber's impulse response. The loss is defined as the discrepancy between the input BGS $y$ and the synthesised spectrum $\hat{y}$, and is minimized via gradient descent. During this iterative optimization, the predicted parameters $y_{pre}$ gradually converge to physically plausible values. Once the optimization reaches convergence, the BFS with high SR predicted by PSRN can be obtained. Moreover, the predicted parameters can be used to reconstruct the high SR BGS through the Lorentzian shape.

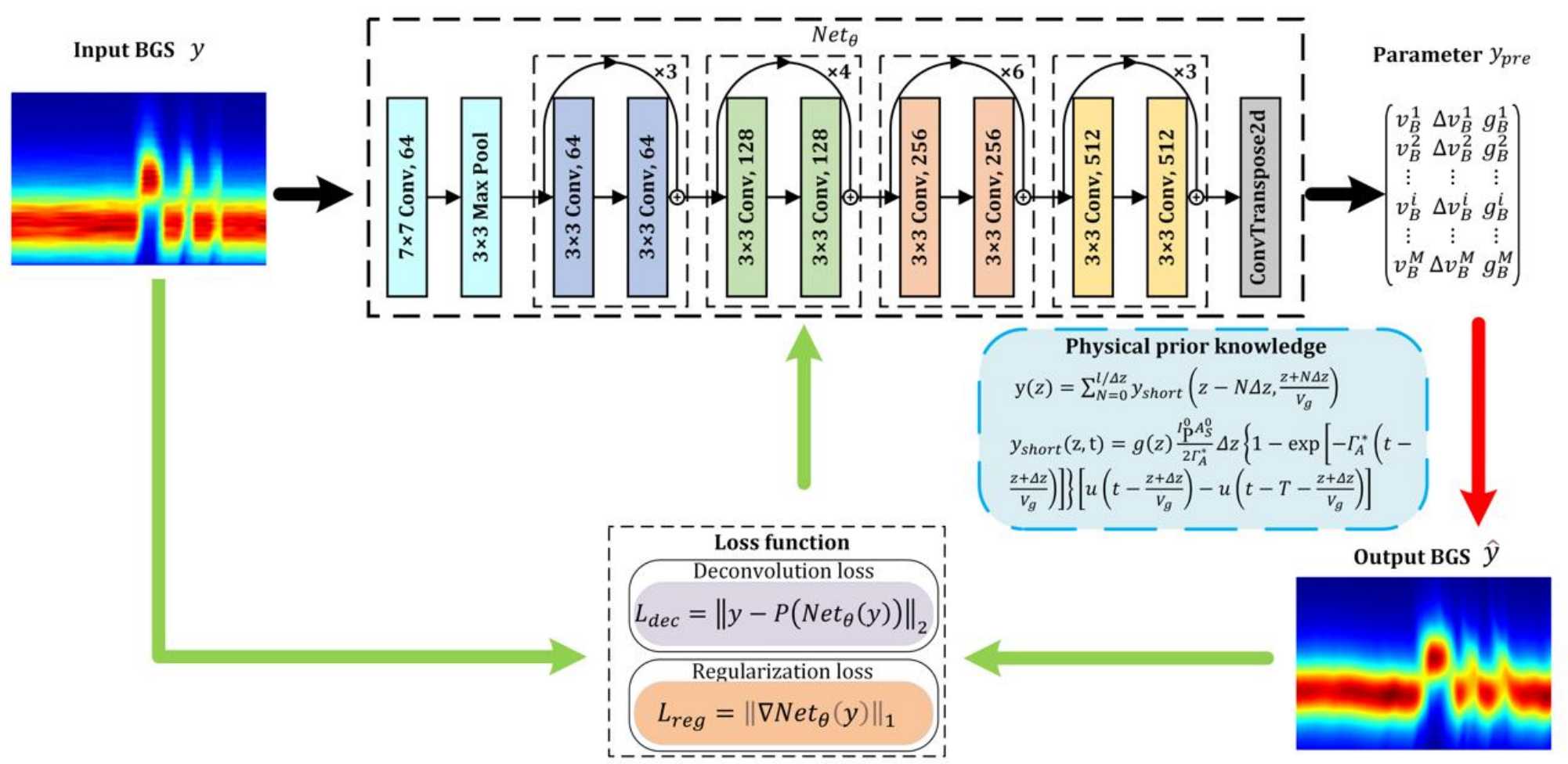

Fig. 1. Schematic illustration of the pipeline of PSRN.

### 2.1 Manually designed network structure

The manually designed network $Net_\theta(\cdot)$ comprises three main components. The first component includes the input layer, a convolutional layer, and a maxpooling layer. The input layer accepts two-dimensional BGS data of size $f_N$ × N, where $f_N$ represents umber of swept frequencies of the BGS (determined by the sweep step), and N denotes the number of BGS traces along the fiber length. The second component is a deep feature extraction module composed of 16 residual blocks, forming a total of 32 convolutional layers. Each residual block contains two 3×3 convolutional layers and a shortcut connection between the input and output, following the standard ResNet design [22]. Based on the number of convolutional kernels, the residual blocks are divided into four types with 64, 128, 256, and 512 kernels, respectively. Among the 16 residual blocks, these four types are distributed as 3, 4, 6, and 3 blocks, respectively. During the process of network iteration, the size of output feature maps gradually decreases in the frequency direction but remains unchanged in the fiber length direction. The third component is an upsampling layer, which is responsible for transforming the intermediate feature maps to the final output dimension. Ultimately, the network produces an output matrix of size 3 × N, representing the predicted BFS $v_B$, Brillouin linewidth $\Delta v_B$, and Brillouin gain coefficient $g_B$ along the fiber.

### 2.2 Physical prior knowledge

Regarding the physical prior knowledge P(·), the time-domain signal measured by BOTDA equipment can be approximated as the convolution between the pulse shape and the system's impulse response, where the spatial resolution is determined by the pump pulse width. The pulse shape itself is defined by the convolution of a short pulse and the Brillouin gain envelope. When a rectangular pump pulse with a width of $T_P$ is used, the Brillouin gain $y_{short}(\mathrm{z,t})$ generated at position z in the fiber over a small uniform fiber segment of length $\Delta z$ can be expressed as the product of the temporal response of the Brillouin gain and the impulse response of the system [23,24]:

$$y_{short}(\mathrm{z,t}) = g(z)\frac{I_P^0 A_S^0}{2\Gamma_A^*}\Delta z\left\{1-\exp\left[-\Gamma_A^*\left(t-\frac{z+\Delta z}{V_g}\right)\right]\right\}\left[u\left(t-\frac{z+\Delta z}{V_g}\right)-u\left(t-T-\frac{z+\Delta z}{V_g}\right)\right] \quad (1)$$

$\Gamma_A = i\pi\left(v_B^2(z)\text{-}v^2\text{-}iv\Delta v_B\right)$ is the frequency detuning parameter, where $\nu_B$ and ν are the BFS at position $z$ and the sweep frequency respectively, and $\Delta\nu_B$ is the intrinsic Brillouin linewidth. The constant $g(z)$ is related to the electrostriction coefficient, The term $I_P^0$ represents the pump pulse intensity, while $A_S^0$ is the intensity of the continuous probe light, $Vg$ is the speed of light in the fiber, $u(\cdot)$ is the Heaviside unit step function, and $T$ is the pump pulse width. The BGS at fiber position $z$ can be solved by concatenating the Brillouin gain of many very short fiber units within the pump pulse width, which can be calculated by [23,24]:

$$\mathrm{y} = \mathrm{P}(z) = \sum_{M=0}^{l/\Delta z} y_{short}\left(z - M\Delta z, \frac{z+M\Delta z}{V_g}\right) \quad (2)$$

where $l$ is the length of pump pulse, Δz is the length of a short fiber unit, and P(·) is the physical prior model.

### *2.3 Loss function*

To ensure the reliability of deconvolution results and mitigate the multisolution issues caused by ill-posed problems, we introduce two key losses: deconvolution loss $L_{dec}$ and regularization loss $L_{reg}$

$$L = L_{dec} + \lambda L_{reg} \quad (3)$$

Deconvolution Loss $L_{dec}$: During the forward process, the input BGS data y is processed by the network to estimate the corresponding physical parameters $y_{pre}$. To ensure the estimated parameters $y_{pre}$ are consistent with the true values of $v_B$, $\Delta v_B$, and $g_B$, the network must be optimized accordingly. Given that the proposed high SR reconstruction framework is trained unsupervised and iteratively, we lack paired BGS data and corresponding reference parameters to guide network optimization. To align the estimated parameters $y_{pred}$ with the observed BGS data, the physical prior model P(·) is employed to map the estimated parameters back to BGS data $\hat{y}$. This reconstructed BGS is then compared to the observed input y, forming the deconvolution loss $L_{rec}$

$$L_{dec} = \left\|y - P\left(Net_\theta(y)\right)\right\|_2 \quad (4)$$

where $\|\cdot\|_2$ represents the L2-norm. In contrast to supervised deep learning deconvolution methods that typically focus on end-to-end mapping, the proposed method incorporates a physical constraint (the BGS convolution model) into the iteration process, effectively reducing potential inconsistencies between the estimated parameters reflectivity and the observed BGS data.

regularization loss $L_{reg}$: Due to the limited bandwidth of $y$ and the presence of noise, optimizing the network $Net_\theta(\cdot)$ solely by minimizing the reconstruction loss $L_{rec}$ can lead to multiple solutions. These solutions may satisfy the physical equations, but do not correspond to the actual state of the fiber. To mitigate this ill-posed problem, we introduce the Total Variation (TV) regularization loss $L_{reg}$ as an additional constraint during parameters estimation [25]. The original TV regularization method targeted image denoising under Gaussian noise [25], nevertheless it has evolved into a more general technique for inverse problems [26] while retaining its edge preserving property [27].

$$L_{reg} = \|\nabla Net_\theta(y)\|_1 \tag{5}$$

where $\|\nabla Net_\theta(y)\|_1 = \sum_{z=1}^{N} \left| y_{pre}(z) - y_{pre}(z-1) \right|$ represents the total variation of solution $y_{pre}$. The $\|\cdot\|_1$ represents the L1-norm, and N denotes the number of BGS traces along the fiber length.

## 3. Experiment and results

PSRN was implemented using Python 3.8.19 and PyTorch 2.4.1. All experiments were conducted on a workstation equipped with an Intel(R) Xeon(R) Gold 6136 CPU, 256 GB of RAM, and an NVIDIA TITAN RTX GPU. The network was trained using the Adam optimizer with a learning rate of $10^{-4}$, which was used to update both the weights and biases during the optimization process. Notably, the weights and biases of the final layer are initialized to 0 and 0.1, respectively. This deliberate initialization strategy stabilizes the network's output during the early iterations and mitigates large prediction deviations that may arise from random initialization. In this study, the input BGS length is flexible and can be adjusted as needed, with the maximum supported size primarily limited by the available GPU memory. The network typically required 2,000 epochs to converge to a high-quality solution.

### *3.1 Simulation results*

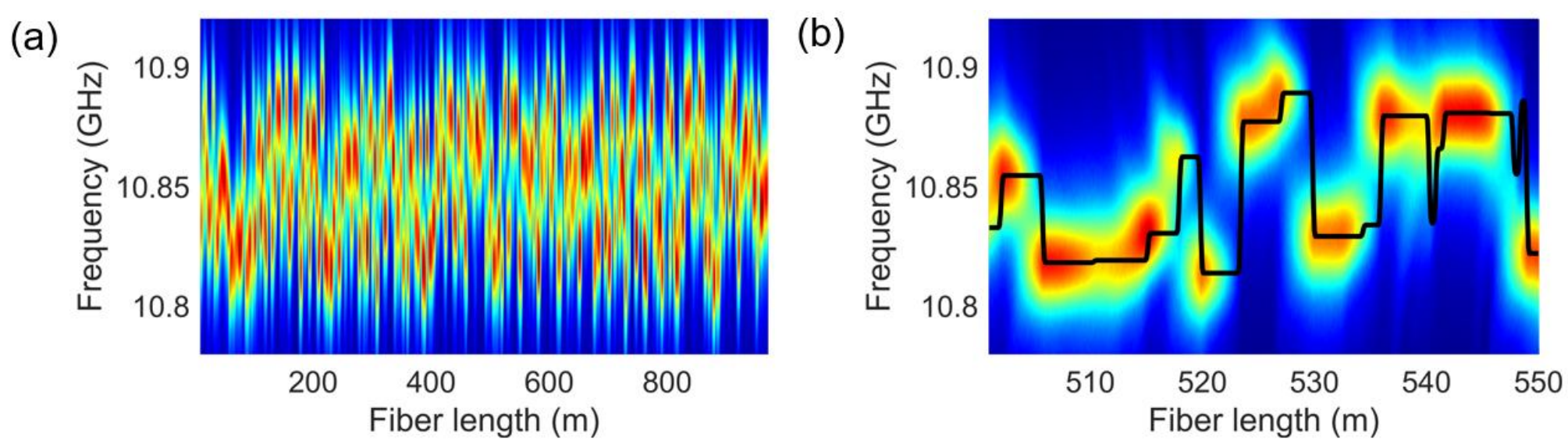

Fig. 2. (a) Simulated 1 km BGS data using a 40 ns pump pulse; (b) BGS at 500-550 m.

A single BGS data of size 71×9800 was generated through simulations, as shown in the Fig. 2(a). The simulation employed a 40 ns pump pulse, corresponding to a theoretical SR of 4 m. With a sampling rate of 1 GSa/s, the simulated 1-km fiber was divided into multiple segments with lengths randomly distributed between 0.5 m and 5 m. The shortest segment length, 0.5 m, represents the minimum detectable scale of BFS variation and thus defines the ideal SR target for this study. Each fiber segment was assigned randomized physical parameters: BFS values ranging from 10.81 to 10.89 GHz, intrinsic Brillouin linewidth between 25 and 35 MHz, and normalized gain intensity between 0.8 and 1.0. The frequency sweep step was set to 2 MHz. It is important to note that the linewidth is the intrinsic Brillouin linewidth, and the simulated BGS linewidth is also related to the pump pulse width, which will be broader when narrow pump pulse is used [28]. To simulate measurement noise, Gaussian white noise with a variance of 0.005 was added to the normalized BGS data, yielding an approximate signal-to-noise ratio (SNR) of 23 dB. Fig. 2(b) shows a view of the 500–550 m region, where the dashed line represents the ground truth BFS. As highlighted by the red box, fiber segments shorter than 4 m show noticeable distortion in the BGS, demonstrating the limitations imposed by the system's SR.

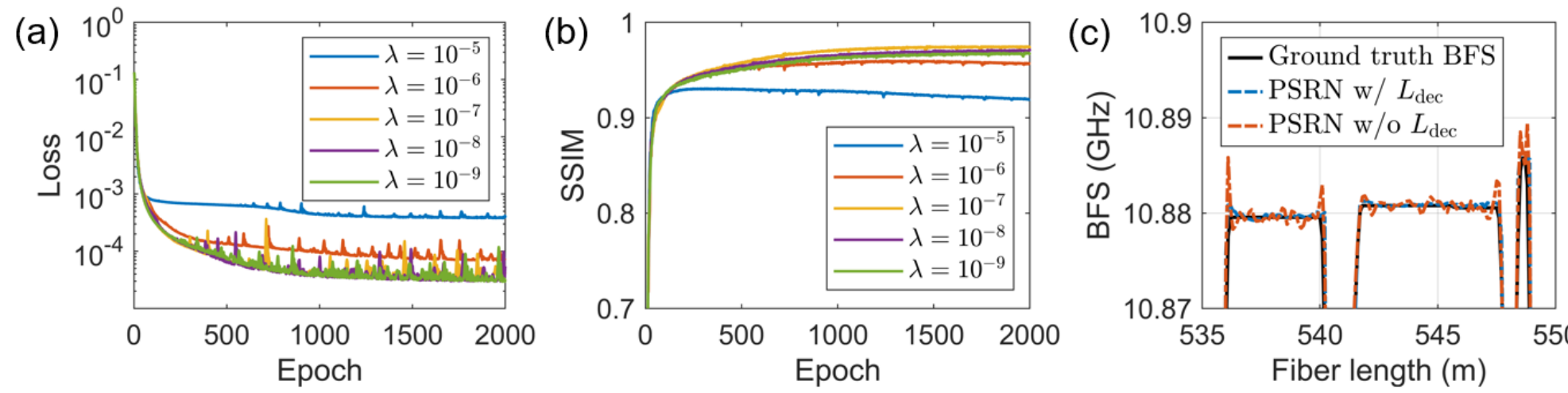

Fig. 3. (a) Evaluation of regularization loss during PSRN-based BGS reconstruction.

To evaluate the impact of regularization loss, we conducted iterative reconstructions of the simulated 1 km BGS using PSRN with manually adjusted regularization weights λ. Fig. 3(a) and 3(b) illustrate the evolution of deconvolution loss $L_{dec}$ and structural similarity index (SSIM) between the reconstructed high-SR BGS and the ground truth BGS over iterations for different values of λ. As λ decreases, the deconvolution performance improves significantly, achieving the best results with a minimum $L_{dec}$ of $3.26\times10^{-5}$ and a maximum SSIM of 0.974 at $\lambda = 1\times10^{-7}$. However, further reducing λ beyond this point does not lead to continued improvement. As shown in the fig.3(c), excessively small λ values weaken the effect of the regularization term, making it ineffective. This leads to the emergence of non-unique solutions, a typical characteristic of ill-posed inverse problems. The BGS deconvolution process in the 500-550 m region is visualized in Fig 4. Specifically, Figs. 4(a) and 4(b) show the input BGS and the corresponding 0.5 m SR ground-truth BGS, respectively. Figs. 4(c)–(h) present the reconstruction results at the 1st, 5th, 10th, 100th, and 2000th epochs. Due to the initialization of the final layer's weights to zero with a bias of 0.1, the output at epoch 1 remains uniform. Nevertheless, through effective integration of physical prior knowledge, PSRN successfully guides the optimization trajectory toward an accurate high-SR reconstruction.

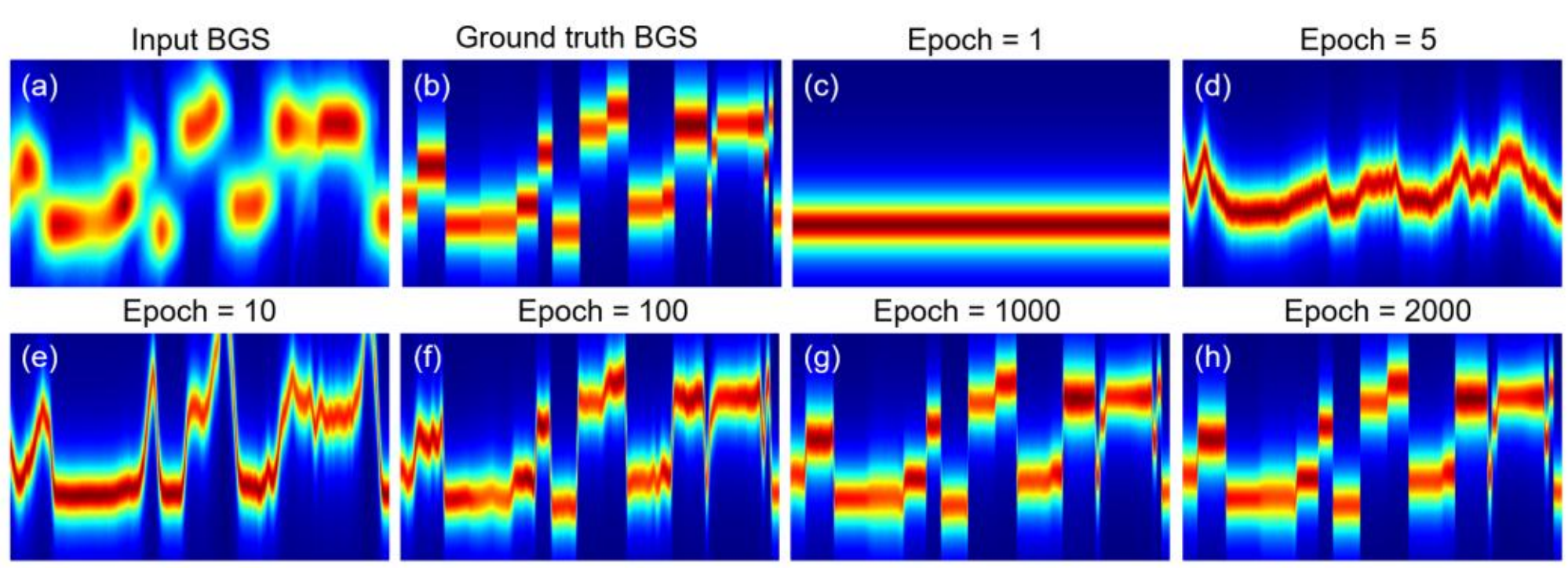

Fig. 4. Visualization of the BGS reconstruction process in 500–550 m region

For high SR BFS extraction, Fig 5 illustrates the optimization trajectory of the model's output BFS during iterations process, where the estimated BFS gradually converges from its initial state to a physically plausible solution. It is evident that PSRN is capable of achieving high-SR reconstructions solely through constraints derived from physical priors, without relying on any labeled data. To evaluate the performance of PSRN, we conducted a direct comparison with a supervised learning approach. Specifically, we employed a STNN adapted from a previous study [21], which accepts fixed-size BGS input of 71×540 and output a 1×540 vector representing the ground truth BFS at 0.5 m SR. The STNN was trained on a synthetic dataset consisting of 10,000 simulated samples, requiring approximately 7.8 hours for data generation and an additional 7 hours for model training. The comparative results are shown in Fig. 6(a), with a detailed view of two consecutive 0.5 m segments shown in Fig. 6(b). In these figures, the black line represents the ground truth BFS at 0.5 m SR, while the yellow line corresponds to the conventional LCF result. The blue and red dashed lines indicate the recovered results by PSRN and STNN, respectively. It is evident that PSRN accurately recovers the BFS, and the recovered result is in great agreement with the ground truth BFS, particularly at the sharp rising/falling edge. This level of rising/falling edge is unattainable for supervised learning methods, which struggle to accurately predict the position of sharp transitions [21]. To quantitatively evaluate performance, we computed the mean absolute error (MAE) relative to the ground truth BFS. PSRN achieved an MAE of only 0.39 MHz, significantly outperforming the STNN, which yielded an MAE of 0.99 MHz.

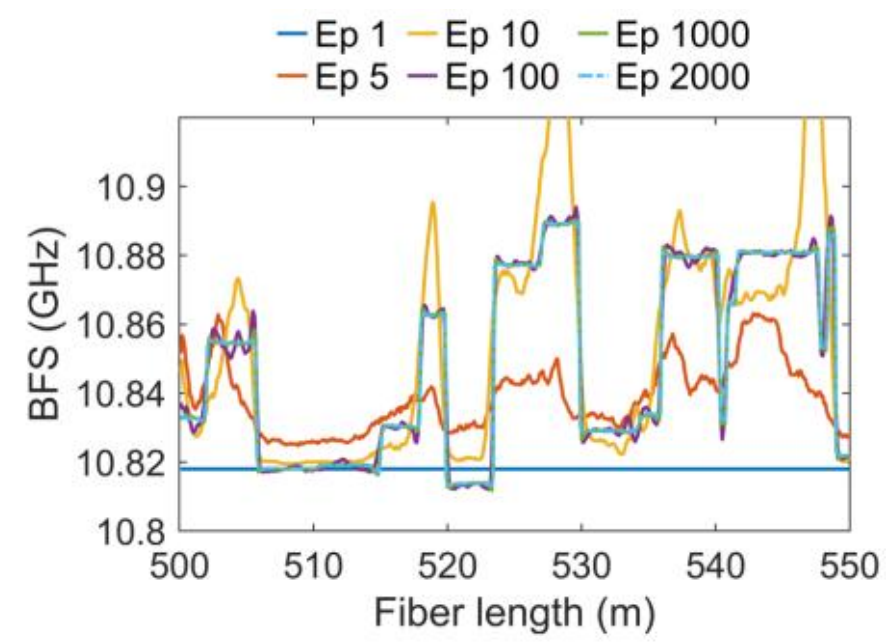

Fig. 5. Optimization trajectory of BFS during PSRN iterations.

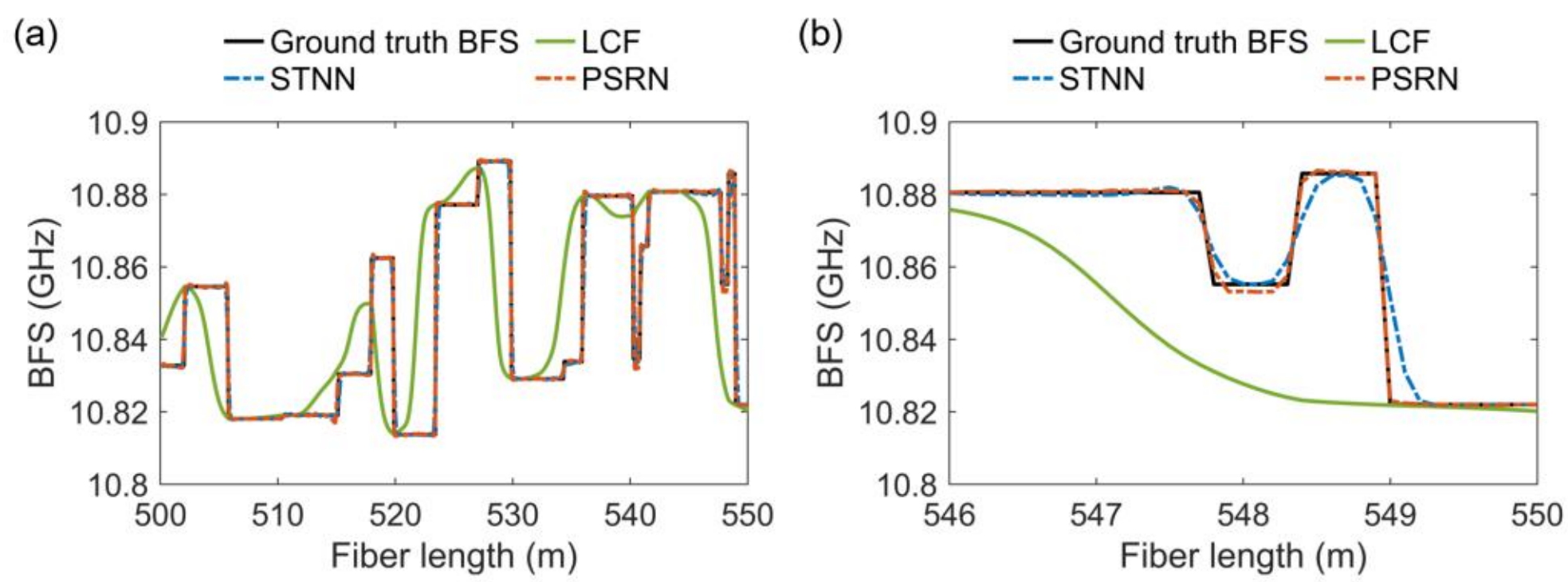

Fig. 6. (a) Optimization trajectory of BFS during PSRN iterations. (b) Comparison of recovered BFS by LCF, STNN, and PSRN. (c) Zoom-in view of two adjacent 0.5 m segments

In addition, PSRN exhibits strong adaptability to varying pulse widths and frequency sweep steps. To validate its generalization capability, we generated simulated BGS using multiple pulse widths (20, 30, 40, 50, and 60 ns) and sweep steps (2, 4, and 8 MHz). Figs 7(a), (d), (g), and (j) show the raw BGS data with a 2 MHz sweep step for pulse widths of 20, 30, 50, and 60 ns, respectively. The corresponding high-SR reconstructions obtained by PSRN are displayed in Figs. 7(b), (e), (h), and (k), and the retrieved BFS results are presented in Figs. 7(c), (f), (i), and (l). PSRN consistently achieves accurate high-SR enhancement across all tested conditions, attaining SSIM values of 0.990, 0.982, 0.974, and 0.961, and corresponding MAEs of 0.21, 0.30, 0.46, and 0.63 MHz for 20, 30, 50, and 60 ns pulse widths, respectively. In contrast, the previously proposed STNN model, trained only on 40 ns data, suffers from poor generalization. When applied to 30 and 50 ns data, STNN exhibits noticeable distortions with increased MAEs of 1.35 MHz and 1.41 MHz. For 20 and 60 ns cases, severe distortions occur, and MAEs rise sharply to 3.09 MHz and 2.85 MHz. Fig. 8(a) summarizes the MAE comparisons between PSRN and STNN, while Fig. 8(b) illustrates PSRN 's robustness across different pulse widths and sweep steps. These results confirm that PSRN is largely invariant to variations in acquisition parameters, thus enabling genuine plug-and-play deployment across diverse BOTDA systems.

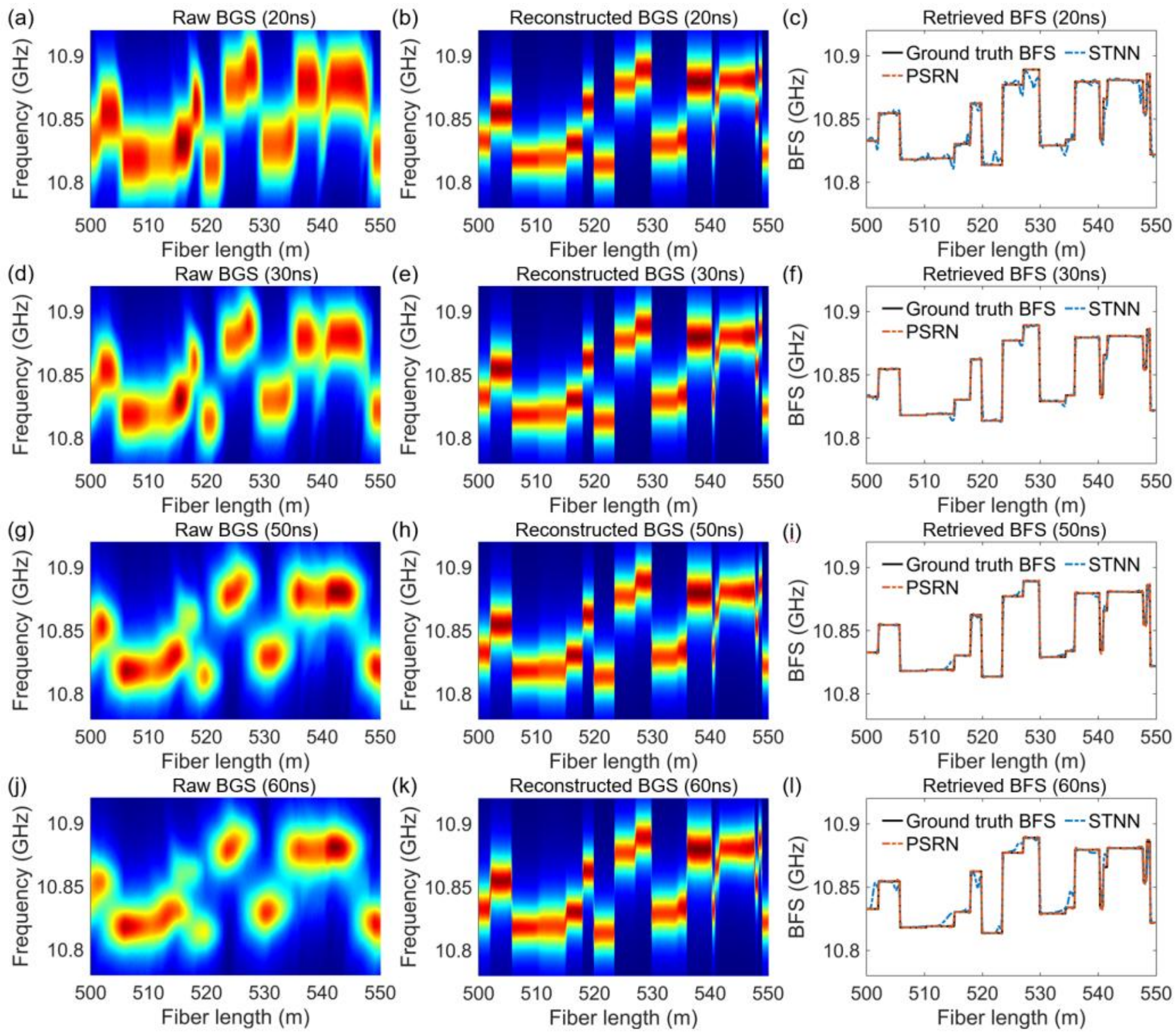

Fig. 7. High-SR BGS reconstruction and BFS retrieval results of PSRN on simulated data with varying pulse widths (20, 30, 50, and 60 ns) at a 2 MHz sweep step. (a)(d)(g)(j): Raw BGS inputs; (b)(e)(h)(k): High-SR BGS reconstructions; (c)(f)(i)(l): Retrieved BFS results.

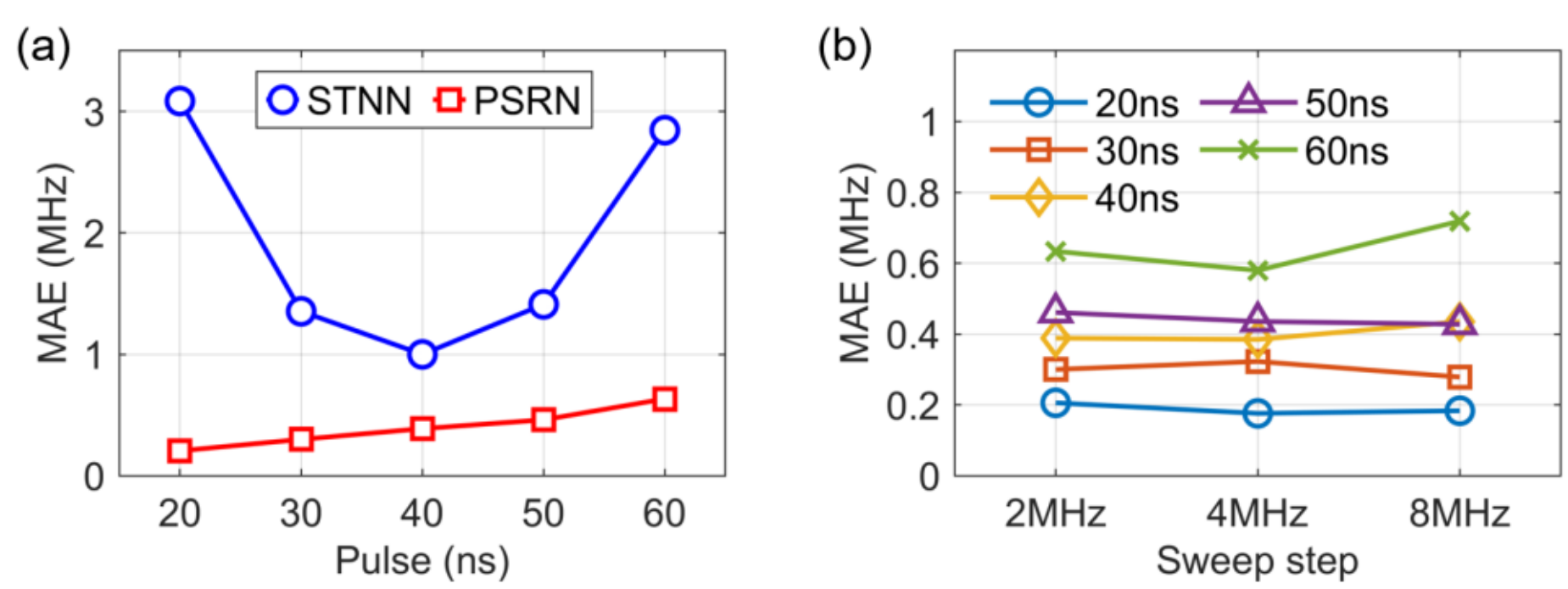

Fig. 8. MAE comparisons under different pulse widths and sweep steps. (a) BFS retrieval MAEs of PSRN vs. STNN for different pulse widths; (b) BFS MAEs of PSRN across various pulse widths and sweep steps.

### *3.2 Experimental results*

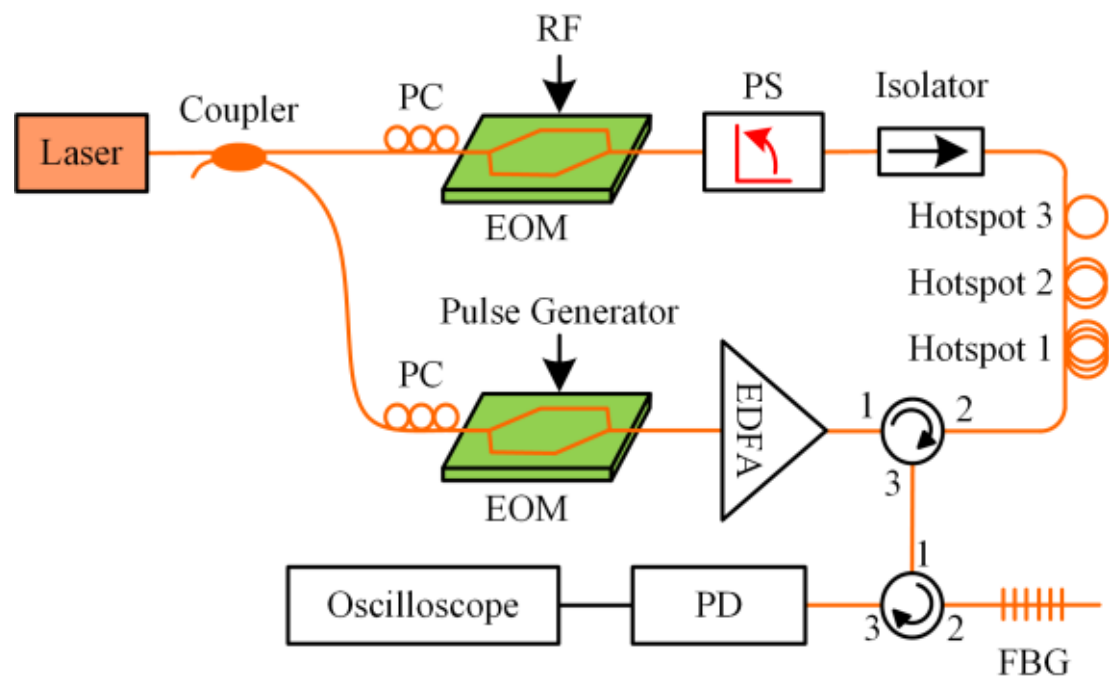

Fig. 9. Experimental setup of the BOTDA system. RF: radio frequency, EOM: electro-optic modulator, PS: polarization switch, EDFA: erbium-doped fiber amplifier, FBG: fiber Bragg grating, PD: photodetector.

To further verify the effect of the CNN, a typical BOTDA sensor as shown in Fig.9 is used to measure the experimental data. The continuous wave light output from the laser source is divided into probe and pump light by a 50:50 optical coupler. The probe light on the upper branch is modulated by an electro-optical modulator (EOM) to sweep the frequency, which is driven by a radio frequency (RF) generator through carrier suppressed double-sideband modulation. And the sweep range is 10.81 GHz to 10.89 GHz in 2 MHz steps. The probe light is finally launched into the 4.9 km long sensing fiber through an isolator. There are three hotspots placed at the end of the fiber, with lengths of 3.3 m, 1 m and 0.5 m, respectively. The lower branch is used for pump light which is modulated by another EOM to generate high extinction ratio pump pulse with fast rising/falling time by using a programmable electrical pulse generator. The erbium-doped fiber amplifier (EDFA) is used to amplify the pump pulse light, and then the amplified pump pulse light passes through a polarization switch (PS) to reduce the polarization fading noise of the Brillouin gain. At the receiver side, a fiber Bragg grating (FBG) is employed to reflect the Brillouin Stokes sideband. The Brillouin signal is finally detected by a photodetector (PD) and then acquired and displayed on an oscilloscope.

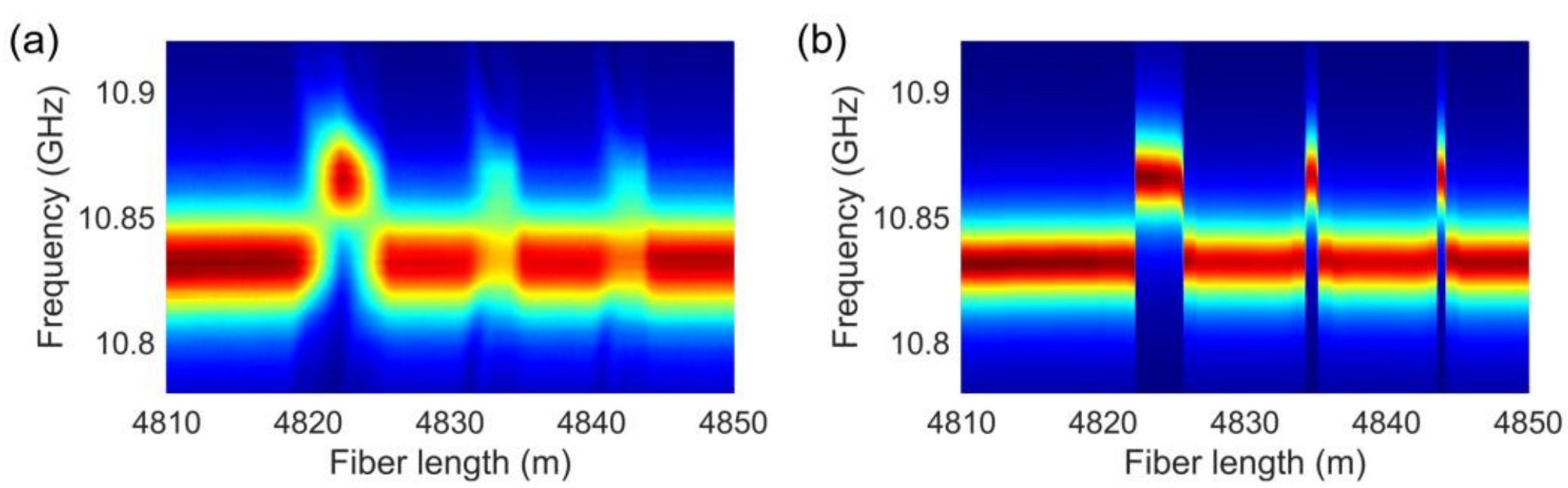

Fig. 10. (a) Raw BGS from the hotspot region. (b) High-SR BGS reconstructed by PSRN. (e) Iterative convergence of PSRN -retrieved BFS.

In the experiment, a 40 ns pump pulse was employed with a sampling rate of 1 GSa/s. The SNR of the raw experimental BGS was about 25.8 dB. The BGS signal from the hotspot region at the fiber end (Fig. 10(a)) was directly fed into PSRN, which successfully reconstructed a high-resolution BGS (Fig. 10(b)). This result clearly shows that PSRN can accurately recover the distorted BGS from hotstops. For high-SR BFS retrieval, we performed a direct comparison among multiple methods. As illustrated by the yellow curve in Fig. 11(a), conventional LCF processing, with a theoretical SR of 4 m, fails to resolve the hotspot accurately. The green solid line and the blue and red dashed lines represent the BFS retrieval results using the 45/40 ns DPP, STNN, and PSRN method, respectively. The average BFS retrieved by DPP and STNN (10.866 GHz) in 3.3 m-long hotspot was used as a reference for calibration, since conventional methods perform reliably at such scales. For the 0.5 m hotspot (Fig. 11(b)), the absolute errors were 1.52 MHz for DPP, 0.34 MHz for STNN, and 0.37 MHz for PSRN. These results demonstrate that PSRN can achieve sub-meter BFS retrieval accuracy directly on experimental data with a 40 ns pulse, without requiring labeled training data or supervised fine-tuning. The retrieved BFS results by PSRN is highly consistent with that of STNN and significantly outperforms the DPP method.

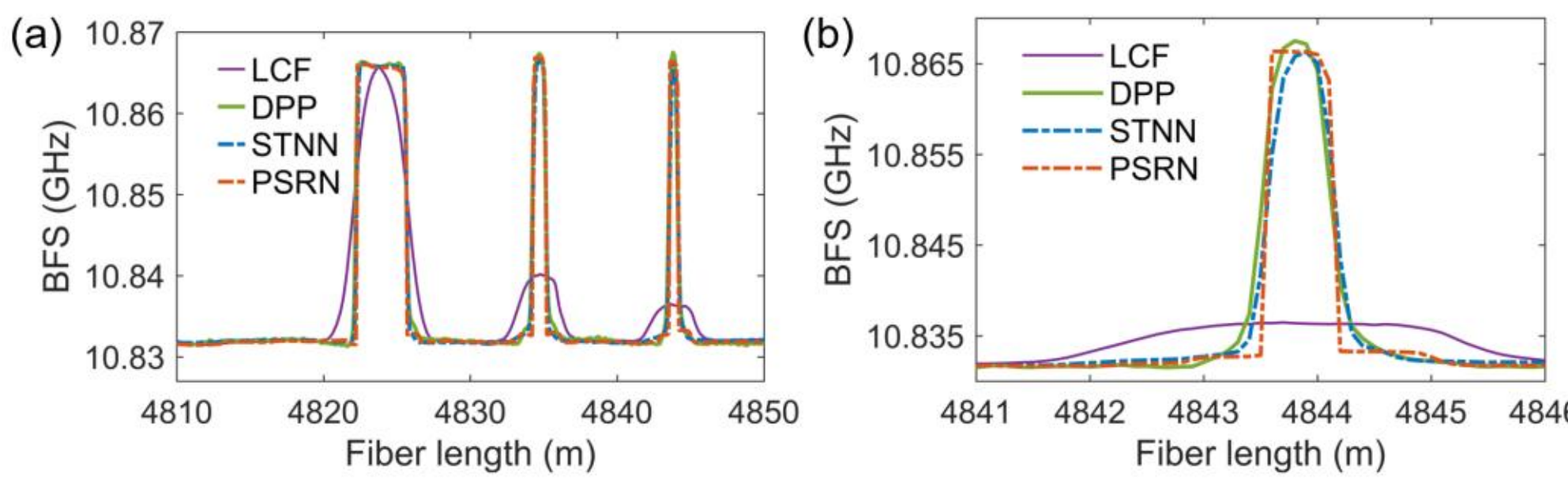

Fig. 11. (a) BFS retrieval comparison among LCF, DPP, STNN, and PSRN methods. (b) BFS retrieval comparison for the 0.5 m hotspot.

To validate the flexibility of the proposed method, experimental BGS signals with varying pulse widths were collected and directly fed into PSRN for iterative reconstruction, without any form of preprocessing. Figs. 12(a), (d), (g), and (j) show the raw experimental BGSs corresponding to 20 ns, 30 ns, 50 ns, and 60 ns pulse widths, respectively. The high-SR reconstructions produced by PSRN are shown in Figs. 12(b), (e), (h), and (k), while the retrieved BFS results are presented in Figs. 12(c), (f), (i), and (l). The results demonstrate that PSRN maintains strong generalization capability on real experimental data, consistently reconstructing high-SR BGSs across all tested pulse widths. From the BFS retrieval results, it is evident that STNN fails to accurately predict high-SR BFS at 20 ns and 60 ns pulse widths, consistent with the trends observed in simulation. For the 0.5 m hotspot, the absolute errors between the PSRN-retrieved BFS and the reference values are 1.86 MHz, 0.59 MHz, 0.30 MHz, and 0.67 MHz for pulse widths of 20 ns, 30 ns, 50 ns, and 60 ns, respectively. In contrast, STNN yields significantly higher errors of 5.01 MHz, 1.25 MHz, 1.03 MHz, and 8.69 MHz, respectively, confirming PSRN's superior adaptability and accuracy under varying measurement conditions.

In addition, PSRN is applicable to experimental data acquired with varying frequency sweep steps. Without any preprocessing, we directly applied iterative reconstruction to experimental BGS measurements containing a 0.5-meter hotspot under sweep steps of 2 MHz, 4 MHz, and 8 MHz. The results are shown in Fig. 13. PSRN successfully retrieved the BFS distributions corresponding to the 0.5-meter hotspot under all sweep conditions, demonstrating strong consistency and robustness. The absolute errors between the retrieved BFS and the calibrated reference values were 0.39 MHz, 0.32 MHz, and 0.31 MHz for 2 MHz, 4 MHz, and 8 MHz sweep steps, respectively.

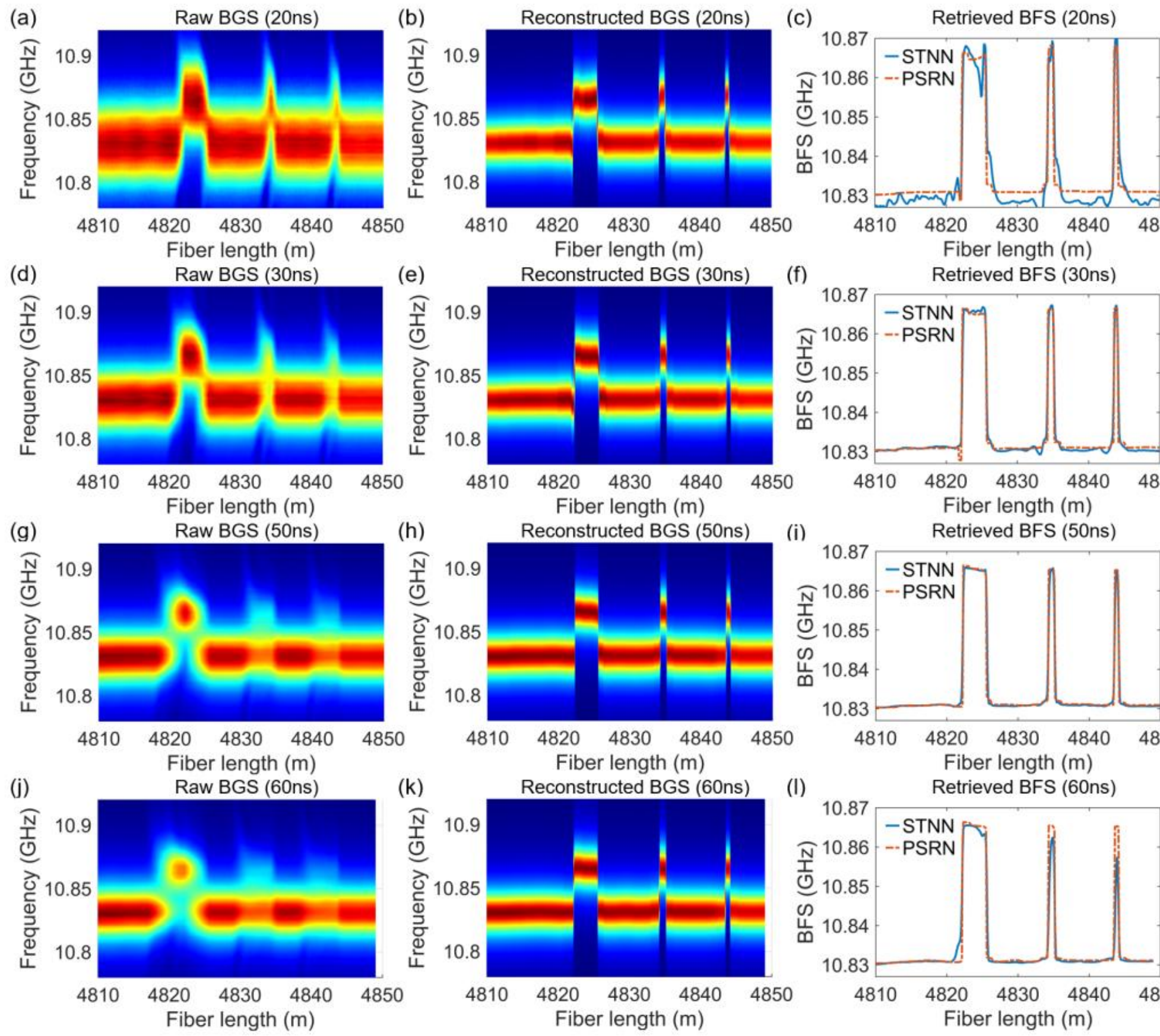

Fig. 12. Experimental BGS reconstruction and BFS retrieval across different pulse widths using PSRN. (a), (d), (g), (j): Raw BGS for 20 ns, 30 ns, 50 ns, and 60 ns; (b), (e), (h), (k): Corresponding high-SR BGS reconstructions; (c), (f), (i), (l): Retrieved BFS results.

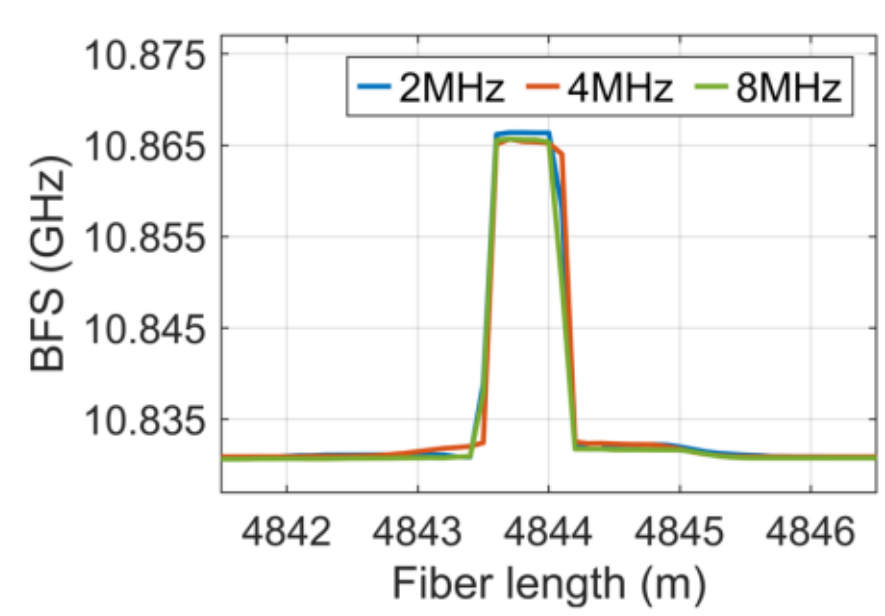

Fig. 13. Experimental BFS retrieval results under different frequency sweep steps using PSRN.

## 4. Conclusion

This study proposes an unsupervised physics-guided deconvolution framework to overcome the inherent SR limitations of BOTDA sensors. The framework enables plug-and-play sub-meter SR BGS reconstruction and BFS retrieval without requiring labeled training data or hardware modifications. The physics-based constraints derived from prior knowledge eliminate the network's dependence on paired low-high SR training data during the training process and enable the model to generalize effectively across different pulse widths and frequency sweep steps. Both simulation and experimental results demonstrate that the proposed method significantly enhances the SR of BOTDA, successfully recovering detailed BGS features and continuous sub-meter variations in the BFS. Compared to conventional SR enhancement approaches, the proposed framework offers stronger nonlinear processing capability, higher prediction accuracy, better flexibility and generalization ability, and superior adaptability across various BOTDA configurations.

**Funding.** National Natural Science Foundation of China (62225110, 61931010); the Major Program (JD) of Hubei Province (2023BAA013); Hubei Provincial Natural Science Foundation of China (2025AFB008)

**Disclosures.** The authors declare no conflicts of interest.

**Data availability.** Both data and codes underlying the results presented in this paper are available in GitHub repository https://github.com/HUST-IOF/BOTDA_PSRN